\documentclass[11pt]{article}

\usepackage{amssymb, amsmath, enumerate, theorem, epsfig}

\begin{document}

\mathchardef\bsurd="1371
\mathchardef\bsolid="132E
\mathchardef\msolid="130E

\def\bb#1{{\bf #1}}
\def\pd#1#2{\frac{\partial #1}{\partial #2}}
\def\ppd#1{\frac{\partial }{\partial #1}}
\def\ppd2#1{\frac{\partial^2 }{\partial #1^2}}
\def\bk{{\bf k}}
\def\bp{{\bf p}}
\def\bv{{\bf v}}
\def\bw{{\bf w}}
\def\bx{{\bf x}}
\def\bz{{\bf z}}

\def\pmb#1{\setbox0=\hbox{#1}%
\kern-.025em\copy0\kern-\wd0
\kern.05em\copy0\kern-\wd0
\kern-.025em\raise.0433em\box0}

\def\bze{{\pmb{$0$}}}
\def\bu{{\pmb{$1$}}}
\def\sm{{\pmb{$\cdot$}}}
\def\vm{{\pmb{$\times$}}}
\def\grad{{\pmb{$\nabla$}}}
\def\div{{\grad\sm}}
\def\curl{{\grad\vm}}
\def\bom{{\pmb{$\omega$}}}
\font\smallrm=cmr8 scaled \magstep 0

\title{\bf Motions in a Bose condensate: X. New results on stability
  of axisymmetric solitary waves of the Gross-Pitaevskii equation}
\author {Natalia G. Berloff${}^1$ and Paul H. Roberts${}^2$\\
${}^1$Department of Applied Mathematics and Theoretical Physics,\\
University of Cambridge, Wilberforce Road, Cambridge, CB3 0WA\\
${}^2$Department of Mathematics, University of California,\\ Los
Angeles, CA, 90095\\ 
}
\date {19 July 2004}
\maketitle
\begin {abstract} The stability of the axisymmetric
solitary waves of the Gross-Pitaevskii (GP) equation is investigated. 
The Implicitly Restarted Arnoldi Method for banded matrices
with shift-invert was used to solve the linearised spectral stability
problem. The rarefaction solitary waves on the upper
branch of the Jones-Roberts dispersion curve are shown to be unstable to
axisymmetric infinitesimal perturbations, whereas the
solitary waves on the lower branch and all two-dimensional solitary
waves are linearly stable. The growth rates of the instabilities on
the upper branch are so small that an arbitrarily specified
initial perturbation of a rarefaction wave at first usually
evolves towards the upper branch as it acoustically radiates
away its excess energy.   This is demonstrated through 
numerical integrations of the GP equation 
starting from an initial state consisting of an unstable rarefaction 
wave and random non-axisymmetric noise. The resulting solution evolves
towards, and remains for a significant time in the vicinity of,
an unperturbed unstable rarefaction wave. It is shown however 
that, ultimately (or for an initial state extremely close to 
the upper branch), the solution evolves onto the lower branch
or is completely dissipated as sound.
  
\end{abstract}

Pacs:{ 03.75.Lm, 05.45.-a,  67.40.Vs, 67.57.De }

\section{Introduction}

Theoretical investigation of the structure, energy, dynamics, and  stability of
vortices in Bose-Einstein condensates (BEC) has received increased attention
since Bose-Einstein condensation was achieved in trapped 
alkali-metal gases at ultralow temperatures; for a comprehensive
review, see for example \cite{fetter}.  Quantised vorticity is  an
  intriguing feature 
of superfluidity, and much effort was therefore devoted to manipulating vortices
and observing their dynamics. This provided a valuable tool for
elucidating the physics of many-particle systems and relating it to 
the quantitative predictions of thermal field theories. The
condensates of alkali vapours are pure and dilute, so that 
the Gross-Pitaevskii (GP) model, which is the so-called `mean-field
limit' of quantum field theories, gives a precise description of their
dynamics at low temperatures. The structure and properties of quantised 
vortices were therefore studied experimentally and then refined, 
both analytically and numerically, by using the GP equation. Conversely, 
the predictions on vortex properties based on the GP model were later
confirmed experimentally. Besides vortices, only one other localised
disturbance has been detected experimentally: the one-dimensional
solitary wave (the `dark soliton') was created from a density
depletion (phase imprinting) in a BEC of sodium atoms \cite{soliton}.

The focus of this report is on a different class of
solitary waves that should exist in a condensate: these are the
vorticity-free axisymmetric disturbances whose existence was
predicted by Jones and Roberts \cite{jr} on the basis of
numerical integrations of
\begin{equation}
-2{\rm i}\, \partial_t \psi =  \nabla^2 \psi +(1- |\psi|^2) \psi,
\label{gp}
\end{equation}
\noindent where $\partial_t=\partial/\partial t$.
In this dimensionless form of the GP equation, the unit of length is 
the healing length, the speed of sound is $c=1/\sqrt{2}$,  
and the density at infinity is $\rho_\infty=|\psi_\infty|^2=1$. 
  In contrast with the vortices and vortex rings that were extensively studied in
  superfluid helium systems long before the experimental realisation
  of Bose-Einstein condensation, the rarefaction solitary waves do not
  exist in superfluid helium as they move faster than the Landau
  critical velocity. Therefore, the existence of the rarefaction solitary waves
  is a unique characteristic of BEC.

\begin{figure}[t]
\caption{(Color online) The JR dispersion curve for the family of the axisymmetric
  solitary wave solutions. The part of the curve that corresponds to
  vortex rings is shown in grey (red). Three rarefaction waves that
  are considered in the text are indicated by circles.
}
\centering
\bigskip
\epsfig{figure=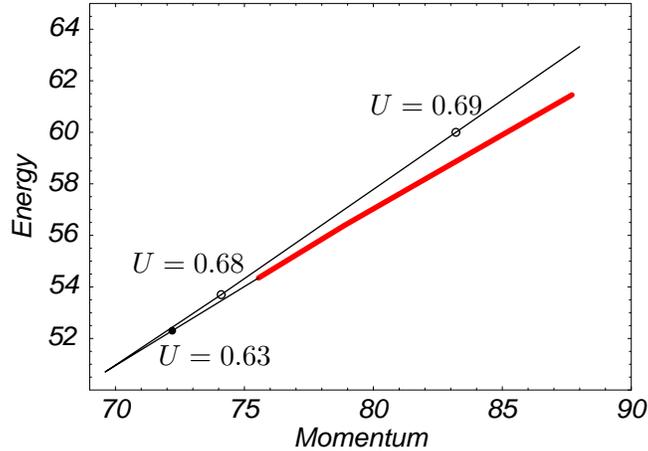, height = 2.4 in}
\begin{picture}(0,0)(0,0)
\put(-110,130) {$U=0.69$}
\put(-200,70) {$U=0.68$}
\put(-190,35) {$U=0.63$}
\end{picture}
\label{cusp}
\end{figure}

Jones and Roberts found all solitary wave solutions of the GP equation
in two dimensions (2D) and three dimensions (3D).
In a momentum energy ($pE$) plot, the 3D sequence,
which we call the `JR dispersion curve', 
has two branches meeting at a cusp where $p$ and $E$ simultaneously
assume their minimum values, $p_c$ and $E_c$. Here
\begin{eqnarray}
p &=& \frac{1}{2{\rm i}}\int[(\psi^*-1)\partial_z\psi
-(\psi-1)\partial_z\psi^*]\,dV\,,\label{pdef}\\
E &=& {\frac{1}{2}}\int|\nabla\psi|^2\,dV\ 
+\ {\frac{1}{4}}\int(1-|\psi|^2)^2dV\,,\label{Edef}
\end{eqnarray}
where $z$ is the direction in which the wave propagates and 
the direction about which it is axisymmetric.
For each $p$ in excess of the minimum $p_c$, two values
of $E$ are possible, and $E \to \infty$ as $p \to \infty$ on each branch. 
On the lower (energy) branch the solutions are asymptotic to 
large circular vortex rings. As $p$ and $E$ decrease from infinity 
on this branch, the solutions begin to lose their similarity to
vortex rings. Eventually, for a momentum $p_0$ slightly in excess 
of $p_c$, they lose their vorticity ($\psi$ loses its zero), and
thereafter the solutions may better be described as
`rarefaction waves'. The upper branch solutions consist 
entirely of these waves and, as $p \to \infty$, they asymptotically
approach the rational soliton solution of the Kadomtsev-Petviashvili
Type I (KPI) equation. Fig.~\ref{cusp} shows the two branches,
the cusp, and the three rarefaction waves that will provide the
principal examples below.

\begin{figure}[t]
\caption{The minimum of the real part of the wavefunction, $Re(\psi)=u$,
  for the JR solitary waves as a function of velocity $U$. The open
  circle marks the point defect; closed circles show computed points
  through which the continuous line is drawn.
}
\centering
\bigskip
\epsfig{figure=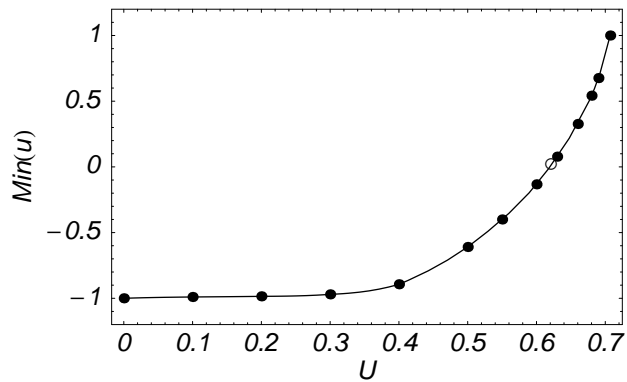, height = 2 in}
\begin{picture}(0,0)(0,0)
\put(-33,85) {\circle{4}}
\end{picture}
\label{min}
\end{figure}

The JR sequence can be uniquely characterised in other ways, for example by
the velocity, $U$, of the wave or by min($u$), the minimum of the real part
of $\psi = u +{\rm i}v$. Fig.~\ref{min} shows min($u$) as a function of $U$ 
for the entire family of JR solutions. In the limit $U \to 0$ of large vortex
rings, min($u$) $\to -1$; in the opposite limit, $U \to c$, of large
rarefaction waves, min($u$) $\to +1$. Between these extremes, the case 
min($u$) = 0 deserves special mention, as it separates the rarefaction
waves, which do not possess vorticity, from the vortex-type solutions
which do. In this case, $\psi$ vanishes at a single point, so that
this solution might appropriately be termed a `point defect';
its velocity is approximately $U \approx 0.62$. The cusp ($U \approx 0.65$)
in the $pE-$plot arises because $U = \partial E/\partial p =
E'(U)/p'(U)$, so that extrema of $p$ are simultaneously extrema of $E$.

It was suggested in \cite{jr} as well as in its more detailed sequel
\cite{jpr86} that every solitary wave on the upper
branch is unstable, since it is energetically favourable for it to
`collapse' onto the lower branch of smaller energy at the same momentum.
A Derrick-type argument \cite{derrick64} was used in which
neighbouring axisymmetric states having the same $p$ as the 
upper branch solution to (1) were shown to have a smaller $E$.
It would follow that the solution is unstable provided
that $p$ and $E$ are the only quantities conserved by (1)
in 3D, but this has never been demonstrated. Moreover,
rarefaction waves on the upper branch of the 
JR dispersion curve are seen in numerical simulations:
they evolve from a density depletion of a condensate \cite{b04} and 
appear during condensate formation from a strongly non-equilibrated Bose
gas \cite{svist}. They may exist for a considerable time before they either 
collapse onto the lower branch of the JR dispersion curve or 
lose their energy to sound waves and disappear.

Evidently there is a paradox. On the one hand, if the assumption
that $p$ and $E$ are the only conserved quantities in 3D is
correct and the upper branch solutions are intrinsically
unstable, how could they even form? On the other hand, 
if the assumption is false and the upper branch solutions 
are stable, why do they ultimately disappear? Is it because 
they are metastable or because they lose energy and momentum 
through collisions with other waves in the system?

The goal of this paper is to resolve this paradox. In Section 2,
we demonstrate that the upper branch solitary waves are 
unstable. In Section 3.1, we show that, nevertheless,
a perturbed solitary wave on the upper branch will generally 
evolve towards an unperturbed state on that branch and remain in its vicinity
for a long period. Eventually, however, as illustrated in
Section 3.2, it collapses onto the lower branch or disappears entirely. 
Another facet of such evolutionary processes is studied in Section 3.3,
where it is shown how two rarefaction waves from the upper branch 
that move in the same direction can create a vortex ring. 
In Section 3.4 we demonstrate how unstable rarefaction waves can form
from the evolution of a density depletion. We conclude in Section 4
with a brief summary of our findings.

\section{Linear stability of the rarefaction solitary waves}

\subsection{The eigenvalue problem}

In a reference frame moving with velocity $U$ in the $z-$direction, 
the GP equation (\ref{gp}) is
\begin{equation}
-2 {\rm i}\,\partial_t\psi + 2 {\rm i} U\,\partial_z\psi 
= \nabla^2\psi  + (1 - |\psi|^2)\psi,
\label{tUgp}
\end{equation}
solutions to which must obey
\begin{equation}
\psi\to 1,\qquad\mbox{for}\qquad r\equiv|\bx|\to\infty\,.
\label{bcpsi}
\end{equation}
The ``basic solution" to (\ref{tUgp}) and (\ref{bcpsi}) is the 
solitary wave, $\psi_0$ $(= u_0 +{\rm i}v_0)$, for which
\begin{eqnarray}
 2 {\rm i} U\,\partial_z\psi_0 &=& \nabla^2\psi_0  
+ (1 - |\psi_0|^2)\psi_0\,,\label{Ugp}\\
\psi_0 &\to& 1,\qquad\mbox{for}\qquad r\to\infty\,.
\label{bcpsi0}
\end{eqnarray}
In this Section, we seek to determine the fate of small
perturbations to $\psi_0$. We write 
\begin{equation} 
\psi(\bx,t) = \psi_0(\bx) + \widehat\psi(\bx,t)\,. 
\label{pert}
\end{equation}
On substituting (\ref{pert}) into (\ref{tUgp}) and 
linearising with respect to $\widehat{\psi}$, we find that
\begin{eqnarray}
-2{\rm i}\partial_t\widehat{\psi}\,+\,2{\rm i}U\partial_z\widehat{\psi}\,&=&\,
\nabla^2\widehat{\psi}\,+\,(1-2|\psi_0|^2)\widehat{\psi}\,-\,
\psi_0^2\widehat{\psi}^*\,,
\label{perteq}\\
\widehat{\psi} &\to& 0,\qquad\mbox{for}\qquad r\to\infty\,,
\label{pertbc}
\end{eqnarray}
where $*$ stands for complex conjugation. Further discussion
of (\ref{pertbc}) is postponed to \S2.4. The objective now is 
the general solution of the initial value problem, i.e., the
solution of (\ref{perteq}) and (\ref{pertbc}), starting from
an arbitrarily specified initial state. Following a well-trodden 
path in stability theory, we suppose that the solution can be
expressed as a linear combination of a complete set of normal
mode solutions. Alternative approaches to the initial value
problem will be described in \S3.

\subsection{Real growth rates}

Equation (\ref{perteq}) admits normal mode solutions in which
\begin{equation}
\widehat{\psi}(\bx,t) = p_1(\bx)\mbox{e}^{\sigma t}\,,\qquad
\widehat{\psi}^*(\bx,t) = p^*_2(\bx)\mbox{e}^{\sigma t}\,,
\label{represent1}
\end{equation}
and
\begin{equation}
Lp_1-\psi_0^2p_2^* = -2{\rm i}\sigma p_1\,,\qquad
L^*p_2^* - \psi_0^*{^2}p_1 =\,\,2{\rm i}\sigma p^*_2\,,
\label{p12eqs}
\end{equation}
where
\begin{equation}
L\,=\, \nabla^2\, +\, 1\, -\, 2|\psi_0|^2\, -\, 2{\rm i}U\partial_z\,.
\label{Ldef}
\end{equation}

If the growth rate $\sigma$ is real, we may take
\begin{equation}
p_1 = p_2 = p = u + {\rm i}v\quad (\mbox{say})\,,
\label{rep1}
\end{equation}
where $u$ and $v$ are real, both being proportional to
$\cos m\phi$ or $\sin m\phi$ where $m$ is an integer
and ($s$, $\phi$, $z$) are cylindrical coordinates.
We shall be primarily interested in the axisymmetric
case $m = 0$. Equations (\ref{p12eqs}) require
\begin{equation}
Lp-\psi_0^2p^* = -2{\rm i}\sigma p\,,
\label{peq}
\end{equation}
or equivalently
\begin{eqnarray}
 \nabla^2 u + 2 U \partial_z v  
+  (1 - 3 u_0^2 -v_0^2)u -2u_0v_0v &=&  \quad 2\sigma v\,,
\label{stability1}\\
 \nabla^2 v - 2 U \partial_z u 
+ (1 -  u_0^2 - 3v_0^2)v -2u_0v_0u &=& - 2\sigma u\,,
\label{stability2}\\
u, v\quad \to 0,\quad \mbox{for}\quad r&\to&\infty\,,
\label{stability3}
\end{eqnarray}
where $\psi_0 = u_0 + {\rm i}v_0$.

The numerical solution of (\ref{stability1}) -- (\ref{stability3}), 
was achieved by first making the change of independent variables
suggested by \cite{jr}: $z'=z$ and $ s'=s\surd(1-2 U^2)$.
We mapped the infinite domain onto the box
$(0,\textstyle{\frac{1}{2}}\pi)\times
(-\textstyle{\frac{1}{2}}\pi,\textstyle{\frac{1}{2}}\pi)$ 
using the transformation
\begin{equation}
\widehat z=\tan^{-1}(C z'),
\label{transf}
\end{equation}
with a similar transformation for $s'$; here $C$ ($\sim 0.4-0.5$) is
a constant chosen at our convenience. We also write $\psi = \Psi + 1$, 
so that the boundary conditions at infinity are $\Psi = 0$.

We used the Newton-Raphson iteration procedure (using a banded matrix
 linear solver based on the bi-conjugate gradient stabilised iterative method with
preconditioning) in the frame of reference
moving with velocity $U$ to find solitary waves from solutions of
\begin{eqnarray}
2{\rm i}U R \cos^2\widehat z\,\partial_{\widehat z}\Psi_0
&=&{\cal L}\Psi_0 - (\Psi_0 +\Psi^*_0 + |\Psi_0|^2)(1+\Psi_0), \nonumber\\
{\rm and}\quad\Psi_0&\to& 0, \quad {\rm as}\quad 
s,\,z\to\infty,\label{U}
\end{eqnarray}
where
\begin{eqnarray}
{\cal L}\!&\!\!=\!\!&\!R^2 (1-2 U^2)\cos^3\widehat s
\left[\cos\widehat s\,\ppd2 {\widehat s}
\!-\!\left(2\sin\widehat s-\frac{1}{\sin\widehat s}\right)\,
\frac{\partial}{\partial \widehat s}\right]\nonumber\\
& + & R^2\cos^3\widehat z\left[\cos\widehat z\,\ppd2 {\widehat z}\,-\,
2\sin\widehat z\,\frac{\partial}{\partial \widehat z}\right]\,. 
\end{eqnarray}
We also used (\ref{transf}) to transform (\ref{stability1}) -- (\ref{stability3}).
The resulting matrix equation was then discretised and gave the
classic eigenvalue problem, $A{\bf x} = \sigma{\bf x}$, for $\sigma$.
Eigenvalues can often be found successively by the Arnoldi/Lanczos process,
in which the extreme eigenvalues are found first and the remaining eigenvalues
are found successively by the `shift and invert' technique, in which the 
original eigenvalue is shifted by $\lambda$ (say), and $A - \lambda I$
is inverted to transform the equation to 
$(A - \lambda I)^{-1}{\bf x} = \mu{\bf x}$, where $\mu=1/(\sigma-\lambda)$,
so that the original eigenvalue is easily recovered as $\mu^{-1}+\lambda$.
We employed the ARPACK \cite{ARPACK}  collection of subroutines, that
uses the Implicitly Shifted QR technique that is suitable for large scale problems.
 The algorithm is designed to
compute a few ($k$) eigenvalues with user specified features such as
those of largest real part or largest magnitude. Storage requirements
are on the order of $n*k$ locations, where $n$ is a number of columns
of the matrix $A$.  A set of Schur basis vectors for the desired $k$-dimensional eigenspace is computed which is numerically orthogonal to working precision.

The basic
solution was found with a maximum resolution of 250$\times$200, making
the
real matrix generated by (\ref{stability1}) -- (\ref{stability2})
of order $10^5\times10^5$. The resolution of the basic
solution was halved to check the accuracy of the eigenvalues found.

We found that $\sigma^2$ is real for all solutions of
(\ref{stability1}) -- (\ref{stability3}). To explain this,
we first observe that we are interested only in perturbations that 
preserve the mass of the state $\psi_0$ in the sense that,
if use (\ref{pert}) to evaluate the mass flux across
any remote surface, ${\rm S}_\infty$, surrounding a volume $V_\infty$
containing and moving with the solitary wave, it is, to 
second order in $\widehat{\psi}$, 
the same as the mass flux implied by $\psi_0$. This implies that
\begin{equation}
\oint_{{\rm S}_\infty}[(\widehat{\psi}^*\grad\widehat{\psi}
-\widehat{\psi}\grad\widehat{\psi}^*) - 2{\rm i}U|\widehat{\psi}|^2
\widehat{\bf{z}}]\,\sm\,{\bf dS}\,=\,0\,.
\label{Lcond}
\end{equation}
This in turn implies that the operator $L$ is hermitian:
\begin{equation}
\int_{V_\infty}\widehat{\psi}L^*\widehat{\psi}^*dV\,=\,
\int_{V_\infty}\widehat{\psi}^*L\widehat{\psi}\,dV\,.
\end{equation}
From this relation and (\ref{peq}), it follows that
\begin{eqnarray}
\int_{V_\infty}\!\!\![|\,Lp|^2\,
-\,\psi_0^2p^*L^*p^*]\,dV\!\!\!&=& \!\!\!
-\!2{\rm i}\sigma\int_{V_\infty}\!\!\!\!pL^*p^*dV
\nonumber\\
& = &\!\!-2{\rm i}\sigma\int_{V_\infty}\!\!\!\!p^*Lp\,dV\,
\!\!=\!\!-\,2{\rm i}\sigma\int_{V_\infty}\!\!\!\!p^*
[\psi_0^2p^*-2{\rm i}\sigma p]\,dV.
\end{eqnarray}
Re-arranging this equation and using (\ref{peq}) again, we find that
\begin{equation}
\int_{V_\infty}\!\!\![|\,Lp|^2\,-|\psi_0|^4|p|^2]\,dV\, 
= \, -4\sigma^2\int_{V_\infty}\!\!\!|p|^2\,dV.
\label{fundam}
\end{equation}
It follows that $\sigma^2$ can only take real values.
The numerical results indicate that the eigenvalue spectrum is infinite
and discrete and (\ref{fundam}) shows that its limit point is
at $\sigma^2 = -\infty$. But only the necessarily finite number of 
positive $\sigma^2$ are relevant to the physical problem;
see \S 2.3.

\begin{figure}
\caption{The density (a) and the phase (b) of the wave function for the
the fastest growing mode of perturbation for rarefaction solitary wave 
moving in the positive $z-$direction with the velocity $U=0.68$.  
(These results were derived by solving the spectral problem
(\ref{stability1}) -- (\ref{stability2}) numerically). 
}
\centering
\bigskip
\hbox{\epsfig{figure=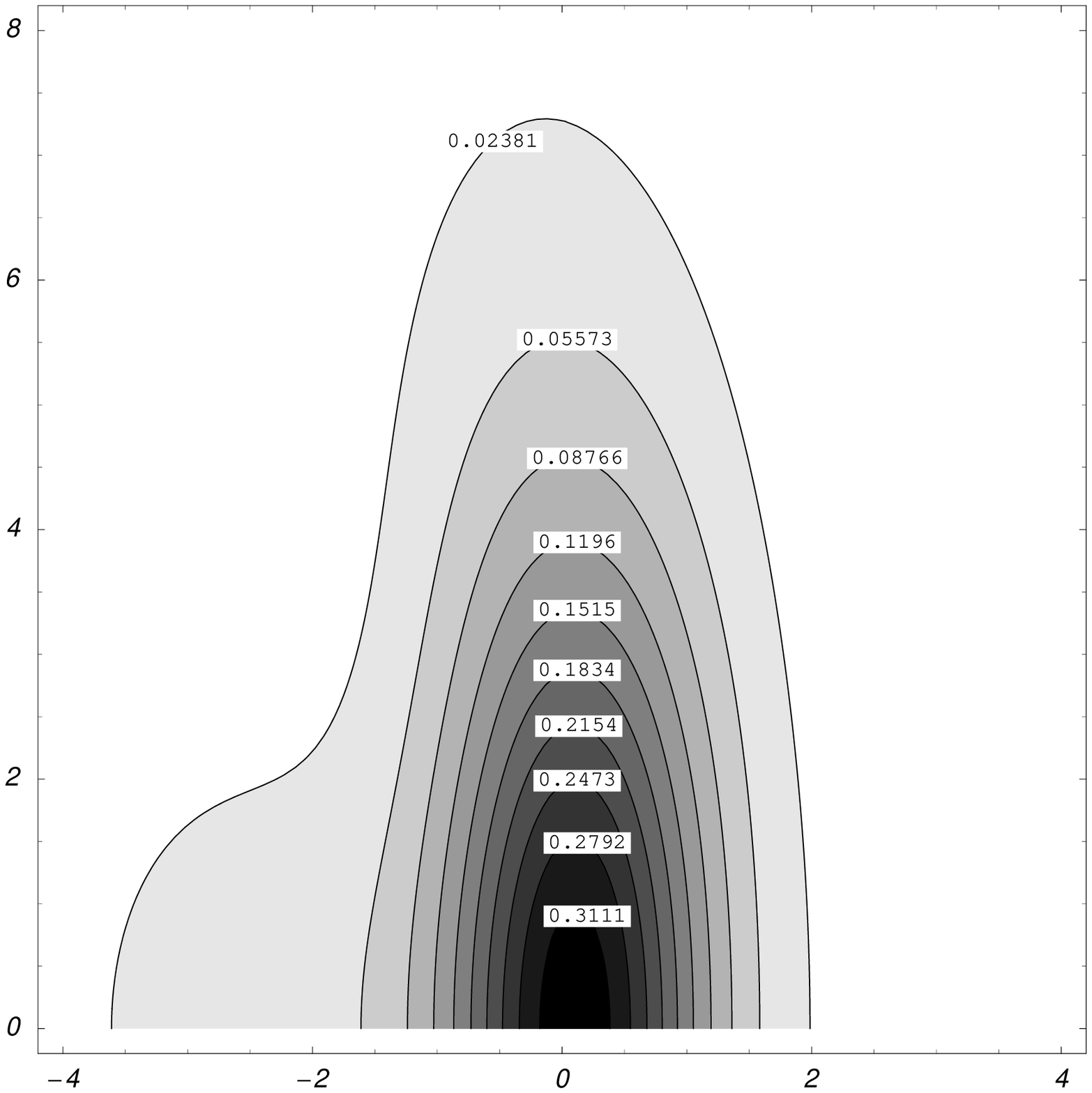, height = 2.3in}\hskip 20 pt \epsfig{figure=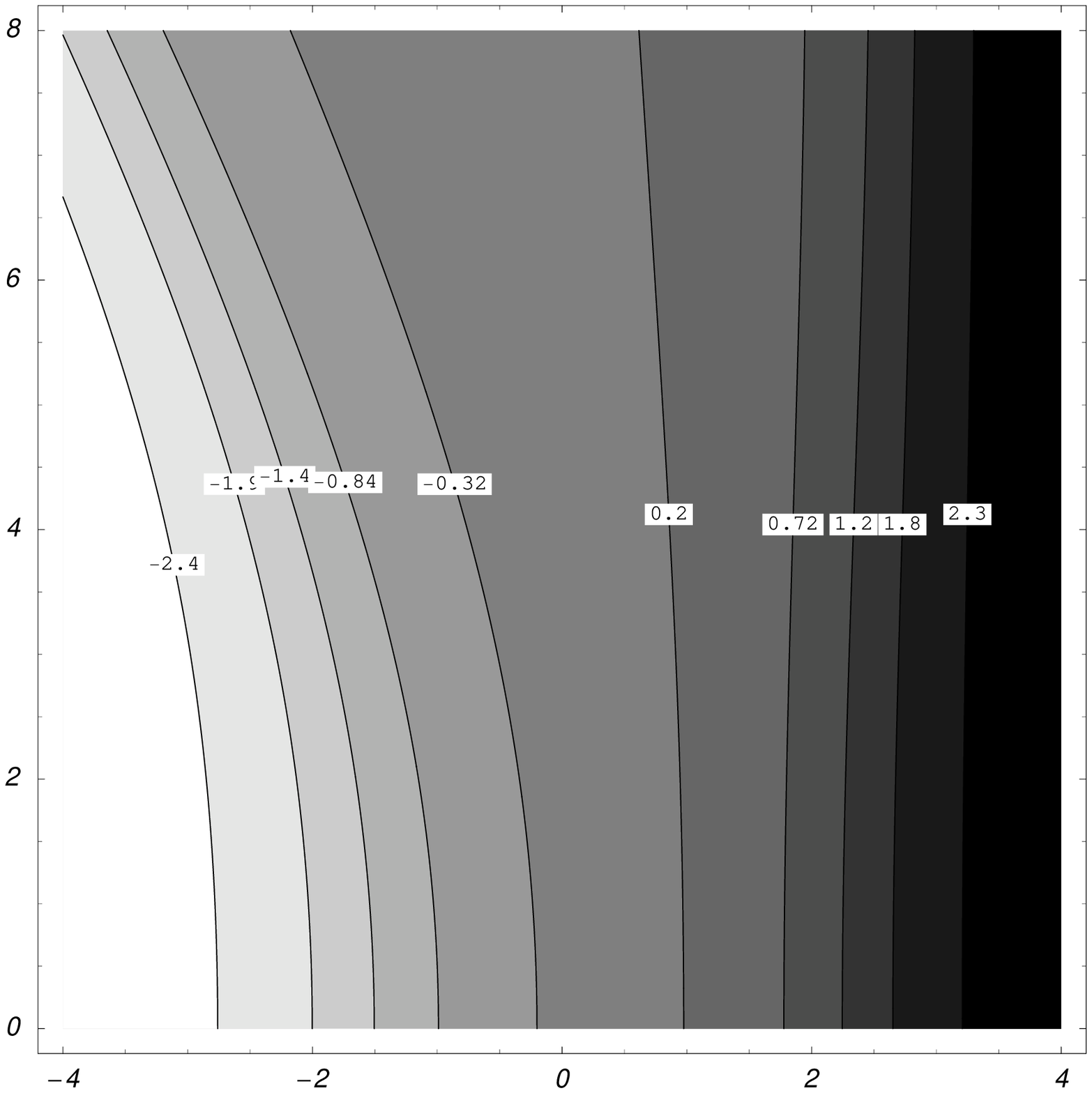, height = 2.3 in}}
\begin{picture}(0,0)(0,0)
\put(20,160) {\Large (b)}
\put(-170,160) {\Large (a)}
\put(55,5) {$z$}
\put(-130,5) {$z$}
\put(-30,110) {$s$}
\put(-216,110) {$s$}
\end{picture}
\label{stab}
\end{figure}

\begin{figure}
\caption{ The maximum growth rate $\sigma_{\rm
  max}(U)$ as a function of the rarefaction wave velocity, $U$.
(These results were derived by solving the spectral problem
(\ref{stability1}) -- (\ref{stability2}) numerically). 
}
\centering
\bigskip
\bigskip
\epsfig{figure=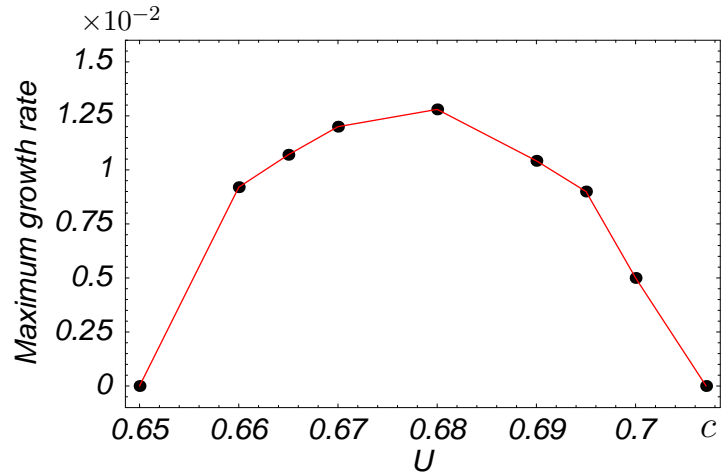, height = 2.3in}
\begin{picture}(0,0)(0,0)
\put(-12,16) {\Large $c$}
\put(-250,170) { $\times 10^{-2}$}
\end{picture}
\label{maxg}
\end{figure}

For the upper branch of the JR dispersion curve, we found only one
family of convincingly positive $\sigma^2$, but this family
(for which $m = 0$) is sufficient to establish that the rarefaction waves 
are unstable to axisymmetric perturbations. The family
seemed to disappear as the cusp $U = U_c$ was approached,
and to become continuous with a family of negative $\sigma^2$ 
modes on the lower branch of the JR dispersion curve. 
Numerical work is incapable of
establishing this fact unequivocally but an analytic argument given
in Appendix A lends support. As $U\to c$, the growth rate of 
the instability tends to zero too (see Appendix B).
One of the more striking results of the numerical work
was the discovery that $\sigma$ is small everywhere on the
upper branch. It was found that $\sigma$ has a single maximum
$\sigma_{\rm max}(U)$, of approximately 0.012, 
attained for $U \sim 0.68$. The fact that $\sigma_{\rm max}(U)$
is so small means that the enfolding time on which the instability 
grows is long, being greater than 1/0.012 $\sim$ 84 in all cases.
 Fig. \ref{stab} shows the density and the phase contour plots
  of the wave function for the
the fastest growing mode of perturbation for rarefaction solitary wave 
moving in the positive $z-$direction with the velocity
$U=0.68$. Fig. \ref{maxg} depicts the maximum growth rate $\sigma_{\rm
  max}(U)$ as a function of the rarefaction wave velocity, $U$.

No instabilities were discovered for the asymmetric ($m \ne 0$) modes
on either branch of the JR dispersion curve, but
the neutral $|m| = 1$ modes, corresponding to infinitesimal 
displacement of the basic state in a direction perpendicular to O$z$,
were located. 

\subsection{Complex growth rates}

When $\sigma$ is complex, the complex conjugate of either of 
(\ref{represent1}) contradicts the other. On making the exchange 
$1 \leftrightarrow 2$ and taking the complex conjugates of both equations 
(\ref{p12eqs}), we see however that, if $\sigma$ is an eigenvalue,
so is $\sigma^*$. Thus solutions to (\ref{perteq}) exist of the
form
\begin{equation}
\widehat{\psi}(\bx,t) = p_1(\bx)\mbox{e}^{\sigma t}
+p_2(\bx)\mbox{e}^{\sigma^*t}\,,\qquad
\widehat{\psi}^*(\bx,t) = p^*_2(\bx)\mbox{e}^{\sigma t}
+p^*_1(\bx)\mbox{e}^{\sigma^*t}\,.
\label{represent2}
\end{equation}

By making use of the Hermitian property of $L$, it follows from
an argument similar to the one that led to (\ref{fundam}) that
\begin{eqnarray}
\int[|Lp_1|^2 - |\psi_0|^4|p_1|^2 - \psi_0^2(p_2^*L^*p_1^*-p_1^*L^*p_2^*)]dV 
\!\!&=& \!\!-4\sigma^2\int|p_1|^2dV\,,\label{iprop1}\\
\int[|Lp_2|^2 - |\psi_0|^4|p_2|^2 - \psi_0^{*2}(p_1Lp_2-p_2Lp_1)]dV 
\!\!&=& \!\!-4\sigma^2\int|p_2|^2dV\,.\label{iprop2}
\end{eqnarray}
On subtracting corresponding sides of these equations we obtain
\begin{eqnarray}
4\sigma^2\int[|p_2|^2-|p_1|^2]dV\!\!&=&\!\!
\int[|Lp_1|^2-|Lp_2|^2 - |\psi_0|^4(|p_1|^2-|p_2|^2)\nonumber\\
&+&\!\!\psi_0^{*2}(p_1Lp_2-p_2Lp_1) 
+ \psi_0^2(p_1^*L^*p_2^*-p_2^*L^*p_1^*)]dV\,.\label{nice}
\end{eqnarray}

In what follows, we shall assume that there are no eigenvalues
(apart from the real $\sigma$ of \S2.1) for which 
\begin{eqnarray}
\int|p_1|^2dV &=& \int|p_2|^2dV\,,\label{eq1}\\ 
\int[|Lp_1|^2-|Lp_2|^2 - |\psi_0|^4(|p_1|^2\!\!&-&\!\!|p_2|^2)\nonumber\\
+ \psi_0^{*2}(p_1Lp_2-p_2Lp_1) 
&+& \psi_0^2(p_1^*L^*p_2^*-p_2^*L^*p_1^*)]dV = 0\,. 
\end{eqnarray}
hold simultaneously. It then follows from (\ref{nice}) that $\sigma^2$ is real.
Therefore the only eigenvalues that are complex are purely imaginary.
The main import of this result is that
\S2.2 has already located all modes that can become unstable.

For modes with imaginary $\sigma = {\rm i}\omega$,
we substitute   
\begin{equation}
p_1 = u_1 + {\rm i}v_1,\qquad p_2 = u_2 + {\rm i}v_2\,,
\label{rep2}
\end{equation}
into (\ref{p12eqs}) to obtain an 8$^{\rm th}$ order system
that replaces the 4$^{\rm th}$ order system
(\ref{stability1}) and (\ref{stability2}):
\begin{eqnarray}
Mu_1 + 2 U \partial_z v_1 - (u_0^2-v_0^2)u_2 -2u_0v_0v_2 
&=&  \quad 2\omega u_1\,,\label{stab1}\\
Mv_1 - 2 U \partial_z u_1 + (u_0^2-v_0^2)v_2 -2u_0v_0u_2 
&=&  \quad 2\omega v_1\,,\label{stab2}\\
Mu_2 + 2 U \partial_z v_2 - (u_0^2-v_0^2)u_1 -2u_0v_0v_1 
&=&  -2\omega u_2\,,\label{stab3}\\
Mv_2 - 2 U \partial_z u_2 + (u_0^2-v_0^2)v_1 -2u_0v_0u_1 
&=&  -2\omega v_2\,,\label{stab4}
\end{eqnarray}
where
\begin{equation}
M = \nabla^2 + 1 - 2u_0^2 - 2v_0^2\,.
\end{equation}

\subsection{Large $r$ behaviour}

It might seem that (\ref{pertbc}) is the obvious and correct
boundary condition to apply to solutions of (\ref{perteq}),
but there are some subtleties. Consider first the solutions
developed in \S2.3.  For $r \to \infty$, $u_1$ and $v_1$
make contributions to $\widehat{\psi}$ that are dominantly 
proportional to
\begin{equation}
r^{-1}\exp[i(k r +\omega t)]\,.
\label{asymptfm}
\end{equation}
Since $u_0 - 1$ = O($r^{-3}$) and
$v_0$ = O($r^{-2}$), the possible values of $k$ implied by
 (\ref{stab1}) -- (\ref{stab4}) are:
\begin{equation}
\omega \pm wk = \pm\textstyle{\frac{1}{2}}k\surd(k^2+2)\,,
\end{equation}
where $w = U z/r$ ($-1/\sqrt{2} < w < 1/\sqrt{2}$). 
The eigenfunction is a 
linear combination therefore of 8 solutions of the form
(\ref{asymptfm}), for the 8 roots of the 2 quartic equations
\begin{equation}
k^4 + 2(1-2w^2)k^2 \pm 8\omega w k -4\omega^2 = 0\,.
\label{quarts}
\end{equation}
It is easily shown that 4 of these roots are real;
2 are positive and 2 negative. These correspond to 
ingoing and outgoing sound waves. They satisfy the
requirement (\ref{pertbc}), but they do not obey
the condition (\ref{Lcond}) needed to establish
that $L$ is self-adjoint. This difficulty can be 
surmounted however by requiring that the waves
are reflected from S$_\infty$. A more serious issue
concerns the remaining 4 complex roots of (\ref{quarts}),
two of which have positive real parts and two negative.
To satisfy (\ref{pertbc}), the eigenfunction must
exclude the two roots for which Im($k$) $<$ 0,
irrespective of whether the remaining $k$ correspond 
to ingoing or outgoing energy flux.

In short, hidden in the succinct statement (\ref{pertbc}),
are physical requirements on the eigenfunctions 
that are far from obvious and may be controversial.
The same is true in the case of real $\sigma$
considered in \S2.2. For this reason, and also to
throw further light on the evolution of perturbations
to the basic state, it was thought desirable to 
carry out the numerical experiments reported in
\S3 below. These attack the initial value problem
without preconceptions such as (10), and without
the necessity of supposing that the perturbation to
$\psi_0$ is infinitesimal. 
\section{Numerical simulations involving rarefaction waves}
In this section we shall report on numerical solutions of the
GP equation (\ref{tUgp}) that show that the solitary waves
on the upper branch can, despite their instability, be surprisingly
robust, but that nevertheless they eventually evolve 
onto the lower branch.
\subsection{Perturbed unstable solitary wave}
Starting with a rarefaction solitary wave from the upper branch of 
the JP dispersion curve, we consider perturbations of the form
\begin{equation}
\psi({\bf x}, t=0) = \psi_0({\bf x}) + {\cal N}({\bf x}),\quad
\mbox{\rm where}\quad {\cal N} =\sum_{\bf k} a_{\bf k}\exp (i{\bf k \cdot x})\,.
\label{noise}
\end{equation}
Here $\cal N$ corresponds to the state of a weakly interacting Bose gas 
in the kinetic regime \cite{kagan}; the phases of the complex amplitudes 
$a_{\bf k}$ are distributed randomly.  To make the disturbance decay 
to zero far from the solitary wave and to give it both axisymmetric and non-axisymmetric components, we take
\begin{eqnarray}
{\cal N}(x,y,z) &=&\frac{1}{M^3} \sum_{{\bf k}
=(l,m,n)}^{M} a_{\bf k}\exp \biggr[i\frac{ \pi}{10}
    \Bigr(lx+\frac{my}{\sqrt{2}} + \frac{nz}{\sqrt{3}}\Bigl)\biggl]\nonumber\\
&\times& \exp[-0.01(x^2 + 2 y^2 + 3 z^2)].
\label{noise2}
\end{eqnarray}
The maximum frequency, $M$,  and the modulus of the complex amplitude
$|a_{\bf k}|$ are two parameters that we vary.

In what follows we say that a perturbed solution evolves
to a JR solitary wave of velocity $U$ if, in a
sufficiently large neighbourhood of its centre and for a significant period
of time, it becomes and remains axisymmetric and form-preserving to good accuracy,  
and if in addition its minimum $u$ corresponds to the $U$ shown in Fig. \ref{min}.
By ``a significant period", we mean an elapsed time exceeding 10,
by ``good accuracy", we mean that the density deviations do not exceed 0.01
and, by ``a sufficiently large neighbourhood", we mean within a sphere
of radius 20 surrounding the origin.

We performed numerical integrations of (\ref{U}) with the 
time derivative $-2{\rm i}\partial_t\Psi$ restored to the
left-hand side.
Since the integration was terminated at a finite time $t$,
before sound waves from the disturbance can reach $r = \infty$,
the boundary condition (\ref{pertbc}) on $\widehat{\psi}$
is irrelevant. We employed the same mapping as in \S2.2. We used fourth order finite differences 
in space  except for second order finite differences at the boundary, and fourth order Runge-Kutta integration in time.  
The initial state at $t = 0$ was given by (\ref{noise}) -- (\ref{noise2}), 
where $\psi_0({\bf x})$ are rarefaction waves from the upper branch 
of the JR dispersion curve. We principally focused on two cases: 
$U=0.68$ and $U=0.69$; see Fig. \ref{cusp}. 

\begin{figure}[t]
\caption{(Colour online) 
Snapshots at $t=0$ (black line), $t=18$ (dark grey or blue
line) and $t=36$ (light grey or red line) of the density of the 
condensate along the direction of the motion. Here $|\psi(0,0,z)|^2$
was obtained by numerically integrating the GP model
(\ref{gp}) for the perturbed rarefaction waves. 
}
\centering
\bigskip
\epsfig{figure=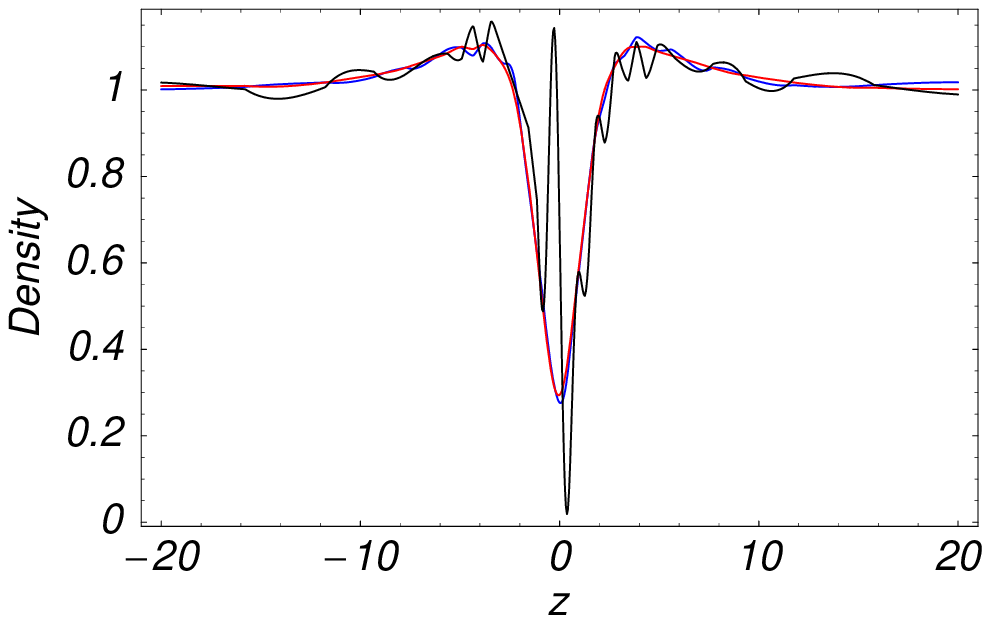, height = 1.8 in} 
\epsfig{figure=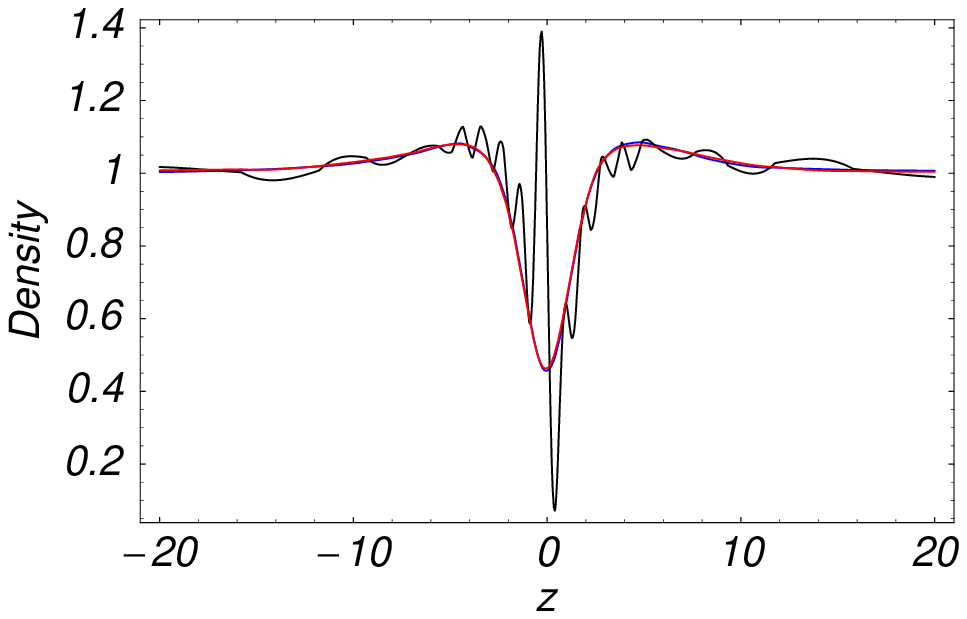, height = 1.8 in} 
\begin{picture}(0,0)(0,0)
\put(-280,60) {\Large $U=0.68$}
\put(-70,60) {\Large $U=0.69$}
\end{picture}
\label{density1}
\end{figure}

In each case, our simulations show that, for $5 \le M \le 20$ 
and $|a_{\bf k}|\le 4$, the wave has, by $t\sim 20$, rid itself 
of the perturbation, apparently by acoustic radiation, and 
has returned to the vicinity of the upper branch, though 
with a velocity that is slightly smaller (by 1\% or less) than
that of the initial $\psi_0$ in (\ref{noise}).
Fig.~\ref{density1} shows snapshots of the evolution of the rarefaction
waves for  $U=0.68$  and $U=0.69$,  for the
initial perturbation (\ref{noise}) -- (\ref{noise2}) with $M=20$ and 
$|a_{\bf k}|=2$. By $t = 18$, each solution had evolved close to an 
unperturbed rarefaction wave, and at $t = 36$ it still remains 
in its vicinity. On a longer time scale ($t > 100$), the rarefaction waves 
either lose their energy and momentum to sound or collapse onto the 
lower branch of the JR dispersion curve.

\subsection{Evolution of unstable rarefaction waves into vortex rings}

In \cite{pade} an algorithm was developed for generalised
rational approximations to the JR solitary waves having the correct 
asymptotic behaviour at infinity. An axisymmetric solitary wave
moving with velocity $U$ along the $z-$axis is well
approximated by  $\psi(s,z)=1+u(s,z) + {\rm i} \,v(s,z)$ where
\begin{eqnarray}
u&=&\frac{a_{00}+a_{10}{z'}^2 + a_{01}s^2
+m c_{20}^{7/4}U(2{z'}^2-f(U)s^2)({z'}^2+f(U)s^2)}
{(1+c_{10}{z'}^2 + c_{01}s^2 + c_{20}({z'}^2+f(U)s^2)^2)^{7/4}},\nonumber\\
v&=&z'\frac{b_{00}+b_{10}{z'}^2 + b_{01}s^2-mc_{20}^{7/4}({z'}^2 + f(U)s^2)^2}
{(1+c_{10}{z'}^2 + c_{01}s^2 + c_{20}({z'}^2+f(U)s^2)^2)^{7/4}},
\label{pade}
\end{eqnarray}
where $z' = z - Ut$ and $f(U) = 1 - 2U^2$. Here $a_{ij}$, $b_{ij}$, $c_{ij}$ 
and the dipole moment $m$ are functions of $U$ determined by series expansions.
For example, the rarefaction wave $U=0.69$ is given by 
\begin{eqnarray}
&\!\!u\!\!&\!=\Bigl(-0.2779\!-\!0.00182{z'}^2\!-\!0.00128\,s^2\!+\!0.00035
(2{z'}^2\!-\!0.0478s^2)({z'}^2\!+\!0.0478s^2)
\Bigr)\!\msolid\!{\cal D}\,, \nonumber\\
&\!\!v\!\!&\!=z'\Bigl(-0.34761\!-\!0.02198{z'}^2\!-\!0.00262s^2\!-\!0.00051
({z'}^2\!+\!0.0478s^2)^2\Bigr)\!\msolid\!{\cal D}\,,
\label{u0.69}
\end{eqnarray} 
where
\begin{equation} 
{\cal D} = \Bigr(1 + 0.11749\,{z'}^2 + 0.01470\,s^2  + 
0.00356({z'}^2 + 0.0478\,s^2)^2\Bigl)^{7/4}\,.
\end{equation} 
By (\ref{pdef}) and (\ref{Edef}), the momentum and energy are
\begin{eqnarray}
p&=&{2\pi}\int (uv_x-vu_x)s\, ds\, dx\,,\label{p3d}\\
{\cal E}&=&\pi\int[u_{z}^2+u_{s}^2+v_{z}^2+v_s^2
-\tfrac{1}{2}(2u+ u^2+v^2)^2]s\,ds\,dx\,,\label{e3d}
\end{eqnarray}
where $u_z = \partial u/\partial z$, $u_s = \partial u/\partial s$, etc.
By substituting the approximation (\ref{u0.69}) into (\ref{p3d}) and (\ref{e3d}), 
we obtain $p\approx 84.15$ and ${\cal E}\approx 61.2$; these may be 
compared with the values $p\approx 83.2$ and ${\cal E}\approx 60.0$ obtained by
from direct integration of (\ref{Ugp}). The maximum residual error, 
calculated as the global maximum of the square of the error amplitude 
[see (29) of \cite{pade}] is $\sim 0.003$. Evidently, (\ref{u0.69})
provides a starting point that is very close to the upper branch.

\begin{figure}[h!t]
\centering
\caption{(Colour online) Snapshots of the 
contour plots of the density cross-section of a condensate, obtained by numerically
integrating the GP model (\ref{gp}). The initial condition was
$\psi=1+u+{\rm i} v$, with $u$ and $v$ given by
(\ref{u0.69}). Black dashed lines show zeros of the real and
imaginary parts of $\psi$ at $t>0$. Their intersection therefore shows the
position of topological zeros.  Both low and high density regions are
shown in darker shades so that regions of intermediate density
are emphasised. Only a portion of an actual computational box is shown.}
\medskip
\vskip 0.2 in
\epsfig{figure=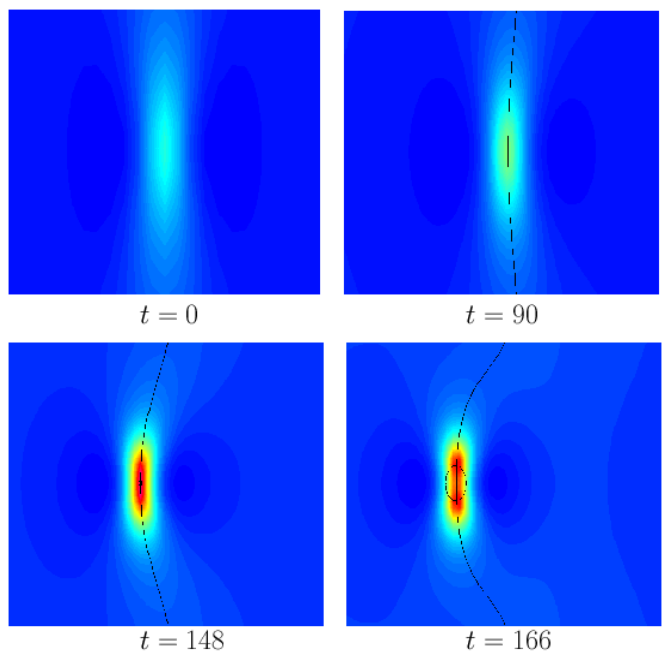,height=3.4in}
\label{contour1}
\end{figure}

To follow the evolution of (\ref{u0.69}) over long times, 
we performed a numerical integration of (\ref{gp}), 
using a finite differences scheme with open boundaries to allow
sound waves to escape (see our previous work \cite{br7} or \cite{br9} 
for a detailed description of the numerical method). The computational 
box has dimensions $D^3 = 100^3$, the space discretization was 
$dx=dy=dz=0.5$, and the time step was $dt=0.1/(dx^2+dy^2+dz^2)$. 
The small initial deviation of (\ref{u0.69}) from the upper branch
grows with time, and the solution slowly evolves into a
vortex ring. Fig.~\ref{contour1} illustrates this process.  
Circulation is acquired at $t\sim 148$ and a
well-formed vortex ring emerges by $t\sim 160$.

\subsection{Vortex nucleation resulting from the interaction of rarefaction waves}

\begin{figure}[h!t]
\centering
\caption{(Colour online) Snapshots of the 
contour plots of the density cross-section of a condensate, obtained by numerically
integrating the GP model (\ref{gp}). The initial condition was
$\psi(t=0)=\psi_0(x-5,s)\psi_0(x+5,s)$, given by (\ref{u0.69}) 
where $\psi_0=1+u+{\rm i} v$. Black dashed lines show zeros of the real and
imaginary parts of $\psi$. Their intersection therefore shows the
position of topological zeros.  Both low and high density regions are
shown in darker shades so that regions of intermediate density
are emphasised. Only a portion of an actual computational box is shown.}
\medskip
\vskip 0.2 in
\epsfig{figure=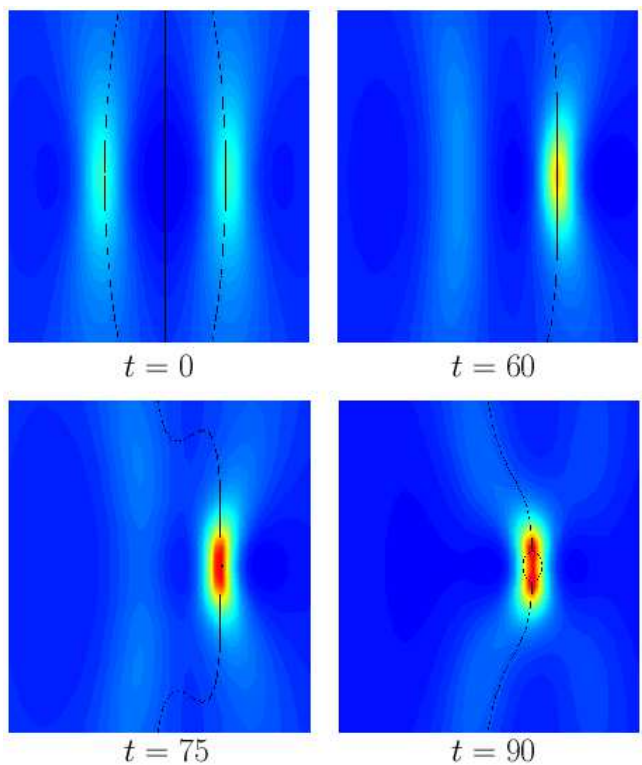,height=4.in}
\label{contour2}
\end{figure}

In \cite{pade} we considered the evolution of two identical
rarefaction waves ($U=0.63$) from the lower branch. These were centred
on the $z-$axis, and moved along it, one behind the other.
It was found that the trailing wave transfered part of its energy 
and momentum to the leading wave, until eventually the latter became 
a vortex ring. We now consider an analogous situation but one in which 
both rarefaction waves initially lie on the upper branch. We numerically
integrated (\ref{gp}) in the frame of reference moving with $U=0.69$, taking  
the initial state to be $\psi(t=0)=\psi_0(x-5,s) \psi_0(x+5,s)$, where
$\psi_0(x,s)$ is defined by (\ref{u0.69}). Fig.~\ref{contour2} shows
snapshots of the subsequent evolution. This is very different from the 
evolution of the two rarefaction waves from the lower branch, 
although there is again a transfer of energy and 
momentum from the trailing wave to the leading wave.
The leading wave does not, however, evolve to a higher energy state 
on the upper branch but instead slowly transforms itself into a vortex 
ring on the lower branch. Apparently, the transfer of energy and 
momentum from the trailing vortex is insufficient to carry the wave 
to a higher energy state on the upper branch; it remains `beneath' that branch 
and can therefore only evolve onto the lower branch; see  Fig.~\ref{cusp}). 

\subsection{Evolution of the density depletions in condensates}

There are several ways of creating rarefaction waves in a condensate, 
that make use of some of the methods described in \cite{b04}. 
A tangle of vortices is created when angular momentum is transmitted
to the condensate by rotationally stirring it with a laser beam 
\cite{caradoc}. In addition to vortices, such stirring creates 
many local depletions in the condensate density. 
It was shown in \cite{b04} that such depletions are unstable and
generate both vortices and rarefaction waves. We shall now show that,
by creating a shallow ellipsoidal depletion in condensate density, 
it is possible to generate rarefaction waves {\it alone}\/ on the 
upper branch of the JR dispersion curve .

\begin{figure}[t]
\caption{(Colour online) 
Snapshots of the contour plots of the density cross-section of a
condensate obtained by numerically integrating the GP model
(\ref{gp}) starting from the spheroidal density depletion (\ref{ic}). 
Black solid lines show zeros imaginary parts of
$\psi$, the straight segment being the axis of symmetry of the
rarefaction wave.  Both low and high density regions are
shown in darker shades so that regions of intermediate density
are emphasised.}
\centering
\bigskip
\epsfig{figure=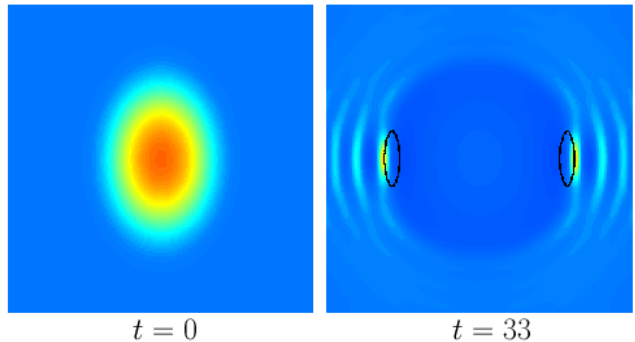, height = 1.8 in}
\label{nucl}
\end{figure}

We numerically integrated (\ref{gp}) taking as our starting
point a spheroidal depletion of condensate:
\begin{equation}
\psi({\bf x},0)=\tfrac{1}{2}+\tfrac{1}{2}
\tanh\left[0.01(z^2+0.5s^2-36)\right].
\label{ic}
\end{equation}
The minimum amplitude of the initial state is $0.33$, at the origin.
After the  condensate fills the cavity, it begins to expand, its
density oscillating around the unperturbed state $\psi=1$. These
depressions in density are unstable, in a manner similar to the
instability of Kadomtsev-Petviashvili 2D solitons in 3D \cite{berloff}. 
This results in the creation of two rarefaction waves moving outwards
in opposite directions. Fig.~\ref{nucl} shows snapshots of the density
in the $z = 0$ cross-section. Well-defined vorticity-free localised 
axisymmetric disturbances -- rarefaction waves --- are formed at around
$t\sim 30$. Their axes of symmetry lie on Im$(\psi)=v({\bf x},t)=0$ and
are plotted as solid lines in the $t=33$ snapshot. 

Various stages in the collapse of
a stationary spherically symmetric bubble were elucidated in
\cite{bubble}, where conditions necessary for vortex nucleation
were also established. A similar analysis applies to the 
evolution of the density depletion (\ref{ic}), for which the density 
is everywhere nonzero initially. In particular, it would be possible to 
calculate the critical radius of a spherical depletion, as a function of 
minimum density, for rarefaction waves to be nucleated.

\section{Conclusions}

We have analysed the linear stability problem for solitary
waves of the GP equation. We have shown that the rarefaction solitary waves
on the upper branch of the JR dispersion curve are unstable to
infinitesimal perturbations and we have calculated the maximum growth rate,
and found it to be small, being approximately $\sigma_{\rm max} =
\sqrt{2}\times0.012c/\xi \approx 70$ ${\rm s}^{-1}$ 
(taking the healing length $\xi = 0.7\mu$m and  the  Bogoliubov speed of
sound $c = 2.8$ mm s$^{-1}$ as in NIST experiments \cite{soliton}).
Our results indicate that the solitary waves on the lower branch and all
two-dimensional solitary waves are linearly stable. By
numerically integrating the GP equation, we studied the properties of
the unstable solitary waves and, in particular, we showed that the 
system tends to remain close to the unstable solution, 
spending a significant amount of time, of the order of
a few multiples of $\tau_{\rm min} = 1/\sigma_{\rm max}$, 
in its vicinity. Ultimately however it collapsed
onto the lower branch or broke up into sound waves. 

These conclusions suggest that rarefaction 
solitary waves could be experimentally detected and and studied 
in Bose condensates alongside the vortex rings.  They can be
created by laser beams and evolve from a local
depletion of the condensate. Rarefaction waves will also appear
as the result of the self-evolution of a strongly nonequilibrated Bose
gas \cite{svist}. Like the vortex rings, they can in principle
be observed in expanding condensates.  
\section{Acknowledgements}
The authors are grateful for support from NSF grant
DMS-0104288. NGB thanks Professor Chris Jones for a useful discussion,
Dr Paul Metcalfe for his help with installing and testing ARPACK, and
Dr Jeroen Molemaker for a useful subroutine.
\section*{Appendix A: Eigenvalues at the cusp}
It was argued in \S2.1 that, for every $U$, the spectrum of $\sigma^2$,
where $\sigma$ is the growth rate of perturbations is real, 
discrete and infinite, with limit point at
$\sigma^2 = - \infty$, i.e., only a finite number of positive $\sigma^2$
can exist for each $m$. The largest of these is the most significant
since, when positive, it gives the largest growth rate for
instability. Our numerical work strongly suggested that this 
$\sigma^2$ is part of the spectrum for the axisymmetric
mode and that it changes sign precisely at the cusp, $U = U_c$, 
being positive for $U > U_c$ and negative for $U < U_c$.
The aim of this Appendix is to re-enforce this conclusion by
showing analytically that this $\sigma^2$ can change sign only
at the cusp. The argument is analogous to one presented
by Kuznetsov and Rasmussen \cite{kr} for 2D solitary waves.

In the axisymmetric case ($m=0$), (\ref{peq}) always possesses
a neutral solution ($\sigma = 0$) corresponding to an
infinitesimal displacement of $\psi_0$ in the $z-$direction;
we write this as
\begin{equation}
p^{(0)} = \partial\psi_0/\partial z\,.
\label{soln0}
\end{equation}
We now seek a neighbouring solution for which $\sigma$ is small,
vanishing when a double root of $\sigma$ exists.
We expand this solution as
\begin{equation}
p = p^{(0)} + \sigma p^{(1)} + \sigma^2 p^{(2)} + \dots\,.
\label{pexpand}
\end{equation}
Substituting (\ref{pexpand}) into (\ref{peq}), we obtain
\begin{eqnarray}
&&Lp^{(0)}-\psi_0^2p^{(0)*}  =  0,\label{pexp0}\\
&&Lp^{(1)}-\psi_0^2p^{(1)*}  =  -2{\rm i}p^{(0)},\label{pexp1}\\
&&Lp^{(2)}-\psi_0^2p^{(2)*}  =  -2{\rm i}p^{(1)}.\label{pexp2}
\end{eqnarray}

The differential of (\ref{peq}) with respect to $U$ and 
(\ref{soln0}) show that the solution to (\ref{pexp1}) is
\begin{equation}
p^{(1)} = -\partial\psi_0/\partial U.
\label{soln1}
\end{equation}
A consistency condition follows from (\ref{pexp2}).
It is easily seen, from (\ref{Ldef}) and (\ref{pexp0})
and the condition (\ref{pertbc}) on $\widehat{\psi}$ 
at infinity, that
\begin{equation}
\int\left[p^{(0)*}(Lp^{(2)}-\psi_0^2p^{(2)*})
+ p^{(0)}(L^*p^{(2)*}-\psi_0^2p^{(2)})\right]
\,dV = 0.
\end{equation}
It now follows from (\ref{pexp2}) that
\begin{equation}
\frac{dp}{dU} \equiv \frac{1}{\rm i}\int\left(
\frac{\partial\psi_0^*}{\partial z}\,\frac{\partial\psi_0}{\partial U} -
\frac{\partial\psi_0}{\partial z}\,\frac{\partial\psi_0^*}{\partial U}\right)
\,dV = 0.\label{dpdU}
\end{equation}
The left-hand equality here is established by differentiating
(\ref{pdef}) with respect to $U$ and carrying out integrations by parts.

It follows from (\ref{dpdU}) that the double zero of $\sigma$ can 
occur only at the cusp, the only point on the JR dispersion curve
where $p'(U)$ is zero.
\section*{Appendix B: The Gross-Pitaevskii equation for $U \to c$}
In multidimensions  the density depletion of the
solitary rarefaction wave as well as the characteristic inverse length
along the axis of propagation tend to zero as the solitary wave
velocity, $U$, approaches the speed of sound, $c$. This allows us to reduce
the GP equation to the Kadomtsev-Petviashvili type I (KPI) equation,
that describes the propagation of acoustic waves of small amplitude
and positive dispersion.

In \cite{jr} such a reduction was obtained as a compatibility
condition of the equations written to $O(\epsilon^4)$ where 
the small parameter $\epsilon \sim \surd(c-U)$. 
In what follows we derive the eigenvalue problem of the
GP equation in the
limit of $U \rightarrow c$ and show that it coincides with the
eigenvalue problem for the KPI equation for an appropriate scaling of
the eigenvalues.

We start by separating the real and imaginary parts of the basic state
as $\psi_0 = u_0 +{\rm i}v_0$ and of the perturbation as 
$\widehat\psi = u + {\rm i}v$; see Section 2. 
The basic solution satisfies (\ref{Ugp}) so that
\begin{equation}
\nabla^2u_0+(1 -u_0^2-v_0^2)u_0 + 2U\partial_z v_0 = 0,\quad
\nabla^2v_0+(1 -u_0^2-v_0^2)v_0 - 2U\partial_z u_0 = 0,
\label{basic}
\end{equation}
while $u$ and $v$ satisfy (\ref{stability1}) -- (\ref{stability2}) 
where, without loss of generality, we take $m=0$. To make
the reduction to the KPI equation we write
\begin{equation}
z\rightarrow z/\epsilon, \quad s\rightarrow s/\epsilon^2,\qquad
\mbox{\rm so that}\qquad
\partial_z\rightarrow \epsilon\partial_z,\quad
\nabla^2\rightarrow \epsilon^2\partial^2z +
\epsilon^4 \nabla^2_{H},
\end{equation}
where $\nabla^2_H = \partial^2_s +s^{-1}\partial_s$.
Next we make the  transformation
\begin{equation}
u_0\rightarrow 1+\epsilon^2 u_0, \quad v_0\rightarrow \epsilon v_0, \quad
u\rightarrow \epsilon u, \quad v\rightarrow v,\quad
U\rightarrow c + \epsilon^2 U, \quad \sigma \rightarrow \epsilon^3\sigma.
\end{equation}
Note that the transformation of the eigenvalue $\sigma$ defines a
slow timescale, the only one that makes the time-dependent GP equation 
consistent with the time-dependent KPI equation; see below.
 
The equations governing the basic solution and the perturbations become
\begin{eqnarray}
&&\partial_z v_0 - \sqrt{2}u_0 -
   \frac{1}{\sqrt{2}}v_0^2 + \frac{\epsilon^2}{\sqrt{2}}\biggl[
   \partial_z^2 u_0-(3u_0+v_0^2)u_0 + 2U\partial_z v_0\biggr] 
   + O(\epsilon^4) = 0, \label{eq5}\\
&&\partial_z^2 v_0 - \sqrt{2}\partial_z u_0 - (2u_0+v_0^2)v_0 +      \epsilon^2\biggl[\nabla^2_H v_0 - u_0^2 v_0 - 2U\partial_z u_0\biggr]
   + O(\epsilon^4) = 0, \label{eq6} \\
&&\partial_z v - \sqrt{2}(u + v_0v) + \frac{\epsilon^2}{\sqrt{2}}\biggl[
   \partial_z^2 u - (6u_0+v_0^2)u -2u_0v_0 v+ 2U\partial_z v - 2\sigma v\biggr] 
  + O(\epsilon^4) =0,\label{eq7} \\
&&\partial_z^2 v - \sqrt{2}\partial_z u-(2u_0+3v_0^2)v - 2 v_0 u
   \nonumber \\
&& \hskip 40 pt  + \epsilon^2\biggl[\nabla^2_H v - u_0^2 v -2u_0 v_0 u- 2U\partial_z u 
  + 2 \sigma u\biggr]+ O(\epsilon^4)=0.\nonumber\\ 
&&\label{eq8}
\end{eqnarray}

Now we expand all functions and the velocity $U$ in powers of
$\epsilon^2$ as $u_0=u_0^0 + \epsilon^2 u_0^1$, etc. At leading order, 
(\ref{eq5}) -- (\ref{eq8}) give
\begin{eqnarray}
&&\partial_z v_0^0 - \sqrt{2} u_0^0 - \frac{1}{\sqrt{2}}v_0^{02}=0, 
   \label{eq9} \\
&&\partial_z^2v_0^0 - \sqrt{2}\partial_z u_0^0 - (2u_0^0+v_0^{02}) v_0^0=0, \label{eq10} \\
&&\partial_z v^0 - \sqrt{2}u^0 - \sqrt{2}v_0^0 v^0 =0, \label{eq11} \\
&&\partial_z^2v^0 -\sqrt{2}\partial_z u^0 - (2u_0^0+3v_0^{02})v^0 
   - 2 v_0^0 u^0 =0. \label{eq12}
\end{eqnarray}

We first consider (\ref{eq9}) -- (\ref{eq10}) and notice that (\ref{eq9})
implies (\ref{eq10}), so that
\begin{equation}
u_0^0=\frac{1}{\sqrt{2}}\partial_z v_0^0 - \frac{1}{2}v_0^{02}.
\label{eq13}
\end{equation} 
We use (\ref{eq13}) to eliminate $u_0^0$ from the leading order 
expressions in square brackets in (\ref{eq5}) and (\ref{eq6}) which we
denote by $R$ and $S$ respectively. After making simplifications we obtain
\begin{eqnarray}
R&\equiv& \partial_z^2u_0^0 - (3u_0^0+v_0^{02})u_0^0 
    + 2U^0\partial_z v_0^0\nonumber \\
&=& \frac{1}{\sqrt{2}}\partial_z^3 v_0^0 - v_0^0\partial_z^2 v_0^0
  - \frac{5}{2}\Bigl(\partial_z v_0^0\Bigr)^2 
  + \sqrt{2}v_0^{02}\partial_z v_0^0 - \frac{1}{4}v_0^{04} 
  + 2 U^0\partial_z v_0^0\,, \label{R} \\
S&\equiv& \nabla^2_H v_0^0 - u_0^{02} v_0^0 - 2 U^0\partial_z u_0^0
   \nonumber\\
&=& \nabla^2_H v_0^0 - \biggl(\frac{1}{\sqrt{2}}\partial_z v_0^0 
   - \frac{1}{2}v_0^{02}\biggr)^2\!\!v_0^0 - 2U^0\biggl(       \frac{1}{\sqrt{2}}\partial_z^2 v_0^0 - v_0^0\partial_z v_0^0\biggr). 
   \label{S}
\end{eqnarray}
Equations (\ref{eq5}) -- (\ref{eq6}) become
\begin{eqnarray}
&&\partial_z v_0^1 - \sqrt{2}u_0^1 - \sqrt{2} v_0^0
v_0^1 = -\frac{1}{\sqrt{2}}R, \label{q1} \\
&&\partial_z^2 v_0^1 - \sqrt{2}\partial_z u_0^1 
- (2u_0^0+3v_0^{02})v_0^1-2 v_0^0u_0^1  = -S. \label{q2}
\end{eqnarray}
The compatibility of (\ref{q1}) and (\ref{q2}) implies
\begin{equation}
S=\frac{1}{\sqrt{2}}\partial_z R + v_0^0 R,
\label{a17}
\end{equation}
which is the KPI equation as derived in \cite{jr}:
\begin{equation}
2\sqrt{2}U^0\partial_z^2 v_0^0 -\nabla^2_Hv_0^0 + \partial_z\biggl[
  \frac{1}{2}\partial_z^3 v_0^0 - \frac{3}{\sqrt{2}}\biggl(
  \partial_z v_0^0\biggr)^2\biggr] = 0\,.
\label{kp}
\end{equation}

Next we consider (\ref{eq7}) and (\ref{eq8}). At the leading order
both (\ref{eq11}) and (\ref{eq12}) require that
\begin{equation}
u^0 = \frac{1}{\sqrt{2}}\partial_z v^0 - v_0^0 v^0.
\label{eq15}
\end{equation}
Note that, if we had assumed a different scaling for $\sigma$,
we would now face an inconsistency. For example, if we had adopted 
$\sigma = O(\epsilon)$, we would have found at this point 
in the argument that $\sigma=0$, i.e., that $\sigma$ is asymptotically 
smaller than the assumed $O(\epsilon)$. 

At the next order beyond (\ref{eq11}) and (\ref{eq12}),
expressions $R'$ and $S'$ analogous to $R$ and $S$ arise:
\begin{eqnarray}
&&R'=\frac{\partial^2u^0}{\partial z^2}-(6u_0^0+v_0^{02})u^0 -2u_0^0v_0^0 v^0
  + 2U^0\partial_z v^0 - 2 \sigma v^0+2v_0^1v^0, \label{RR} \\
&&S'=\nabla^2_H v^0-u_0^{02} v^0 -2 u_0^0 v_0^0 u^0- 2 U^0\partial_z u^0  
 + 2\sigma u^0 +(2 u_0^1 + 6 v_0^0v_0^1)v^0 + 2 v_0^1u^0. \label{SS}
\end{eqnarray}
We use (\ref{eq13}) and (\ref{eq15}) to eliminate $u_0^0$ and $u^0$
from (\ref{RR}) and (\ref{SS}). The equations (\ref{eq11}) and
(\ref{eq12}) then have the form
\begin{eqnarray}
&&\partial_z v^1 - \sqrt{2}(v^1 + v_0^0v^1) 
  = - \frac{1}{\sqrt{2}}R', \label{q3} \\
&&\partial_z^2 v^1 - \sqrt{2}\partial_z u^1 - (2u_0^0+3 v_0^{02})v^1
  -2 v_0^0 u^1 = -S', \label{q4}
\end{eqnarray}
and the compatibility of (\ref{q3}) and (\ref{q4}) leads to the
eigenvalue problem
\begin{equation}
S'=\frac{1}{\sqrt{2}}\partial_z R' + v_0^0 R',
\label{ev}
\end{equation}
or, after simplification,
\begin{equation}
2\sqrt{2}U^0\partial_z^2 v^1 - 3\sqrt{2}\partial_z[
  \partial_z v_0^0\partial_z v^1] + \frac{1}{2}\partial_z^4 v^1 
  - \nabla^2_Hv^1 = 2\sqrt{2}\sigma \partial_z v^1.
\label{ev2}
\end{equation}
Notice, the spectral problem (\ref{ev2}) is the same one as 
obtained from  (\ref{kp}) written in time-dependent form as
\begin{equation}
\partial_z\biggl[2\sqrt{2}\partial_t V - 2\sqrt{2}U^0\partial_z V +
    \frac{3}{\sqrt{2}}(\partial_z V)^2
    - \frac{1}{2}\partial_z^3 V\biggr] + \nabla^2_H V = 0,
\end{equation}
where $V=v_0^0 + v^1 \exp(\sigma t)$. This proves that the KPI equation and the
GP equation share the same linear stability properties, the eigenvalues
being related by $\sigma_{KP}(c-U_{GP})^{3/2}=\sigma_{GP}$ 
as $U_{GP}\rightarrow c$.

\end{document}